\documentclass{WileyMSP-template}
\usepackage{amsmath,amssymb,amsfonts}
\usepackage{enumitem}
\usepackage[symbol]{footmisc}
\usepackage[sort&compress]{natbib}
\usepackage{url}
\def\jcm{J~cm$^{-2}$}
\def\um{\textmu m}
\def\np{\vspace{4mm}}
\begin{document}

\pagestyle{fancy}
\rhead{}

\title{Generating unconventional spin-orbit torques with patterned phase gradients in tungsten thin films}

\maketitle

\author{Lauren J. Riddiford*}
\author{Anne Flechsig}
\author{Shilei Ding}
\author{Emir Karadza}
\author{Niklas Kercher}
\author{Tobias Goldenberger}
\author{Elisabeth M\"{u}ller}
\author{Pietro Gambardella}
\author{Laura J. Heyderman*}
\author{Ale\v{s} Hrabec*}

\dedication{}

\begin{affiliations}
Dr. L.~J. Riddiford, A. Flechsig, Prof. L.~J. Heyderman, Dr. A. Hrabec\\
Laboratory for Mesoscopic Systems, Department of Materials, ETH Zurich, 8093 Zurich, Switzerland.\\
PSI Center for Neutron and Muon Sciences, Forschungsstrasse 111, 5232 Villigen PSI, Switzerland.\\
Email Addresses: lauren.riddiford@psi.ch, laura.heyderman@psi.ch, ales.hrabec@psi.ch \\
\np
Dr. S. Ding, E. Karadza, N. Kercher, T. Goldenberger, Prof. P. Gambardella\\
Laboratory for Magnetism and Interface Physics, Department of Materials, ETH Zurich, 8093 Zurich, Switzerland.\\
\np
Dr. E. M\"{u}ller \\
PSI Center for Life Sciences, Forschungsstrasse 111, 5232 Villigen PSI, Switzerland.

\end{affiliations}


\begin{abstract}

A key aim in spintronics is to achieve current-induced magnetization switching via spin-orbit torques without external magnetic fields. For this, the focus of recent work has been on introducing controlled lateral gradients across ferromagnet/heavy-metal devices, giving variations in thickness, composition, or interface quality. However, the small gradients achievable with common growth techniques limit both the impact of this approach and understanding of the underlying physical mechanisms. Here, spin-orbit torques are patterned on a mesoscopic length scale in tungsten thin films using direct-write laser annealing. Through transmission electron microscopy, resistivity, and second harmonic measurements, the continuous transformation of the crystalline phase of W films from the highly spin-orbit coupled, high resistivity $\beta$ phase to the minimally spin-orbit coupled, low resistivity $\alpha$ phase is tracked with increasing laser fluence. Gradients with different steepness are patterned in the tungsten phase to create spin-orbit torque channels and, when interfaced with CoFeB, tungsten wires with a sufficiently strong gradient can switch the magnetization without an applied magnetic field. Therefore, exploiting the unique microstructure of mixed-phase W allows precise control of the local electronic current density and direction, as well as local spin-orbit torque efficiency, providing a new avenue for the design of efficient spintronic devices. 
\end{abstract}

\section{Introduction}
To meet the growing demand for energy-efficient data retention, there has been extensive research into optimizing magnetic materials for use as memory elements. Spin-orbit torque-based magnetic random access memory (SOT-MRAM) is an attractive option for reading and writing of magnetic elements at high speed and with long endurance \cite{manchon2019current, grimaldi2020single, krizakova2022spin, nguyen2024recent}. However, these devices are limited by the requirement of an applied magnetic field for current-induced switching, which impedes implementation into industrial devices \cite{krizakova2022spin, miron2011perpendicular}. Thus, there have been many studies on engineering magnetic devices to switch without the need for an applied magnetic field \cite{fukami2016magnetization, van2016field, oh2016field, wang2018field, shi2019all, krizakova2020field, liu2021symmetry, wu2022field, xue2023field, yang2024field, li2025fully, kang2025field}. For deterministic field-free SOT switching, the mirror symmetry of the device along the current direction must be broken. In sputter-deposited, polycrystalline multilayers, this has been achieved through lateral gradients made by oblique deposition \cite{yu2014switching, chuang2019cr}, vertical composition gradients \cite{xie2021controllable, zheng2021field}, tilted magnetic anisotropy \cite{you2015switching, kim2022field}, non-uniform currents \cite{kateel2023field, liu2024efficient}, applied electric fields \cite{cai2017electric, kang2021electric}, and selective damage through ion irradiation or laser ablation \cite{cao2020deterministic, lee2023field}. However, these techniques are difficult to implement on the industrial scale because they either provide limited scalability or are not suitable. \np 

One route forward to provide an industry-compatible approach is to locally tune the spin-orbit torque efficiency in the film plane, although relatively little attention has been paid to exploit this opportunity up to now \cite{lavrijsen2012asymmetric, lee2023position, zeng2025deterministic}. This can be achieved by leveraging metastable materials, in which the metastable crystalline phase has a different spin-orbit coupling than the stable crystalline phase, thus providing a natural platform to precisely and locally tune spin-orbit torque efficiency. Thin films of tungsten, with both a metastable and a stable phase, offer this tunability. In addition, this material is widely applicable and has been investigated over decades for applications in superconductivity \cite{kammerer1965superconductivity, ma2025doping}, solar energy \cite{vijaya2018development}, microelectronics \cite{ganesh2020tungsten}, and spintronics \cite{pai2012spin, hao2015giant, demasius2016enhanced, vudya2021optimization, lu2024enlarged}. $\alpha$-W is the stable, body-centered cubic crystal phase of the material, characterized by low resistivity ($\sim30$~$\mu\Omega$-cm) and desirable electronic properties to replace Cu in ultra-narrow wires \cite{choi2012electron}. Meanwhile, the metastable $\beta$-W phase with A15 crystal structure has been employed to exert spin-orbit torques on magnetic thin films due to its high spin-orbit coupling, which yields a high charge-to-spin conversion efficiency of $\sim0.3-0.5$ \cite{pai2012spin}. Despite the high resistivity of the $\beta$-W phase ($\sim240$~$\mu\Omega$-cm), this is a material of choice for state-of-the-art SOT-MRAM devices \cite{grimaldi2020single, nguyen2024recent, garello2018sot}. Due to the distinct properties of each of these two crystalline phases, there has been extensive work carried out to determine how to stabilize each phase and transform between the two \cite{barmak2017transformation, wang2019demand, sriram2023structural}. However, efforts into understanding phase transformations in W has focused on global, rather than local, transformations, and local control of the W crystallographic phase has not been shown. With recent work demonstrating the use of direct-write laser annealing (DWLA) to locally enact heat-induced transformation of physical properties \cite{pinheiro2024direct, giacco2024patterning, riddiford2024grayscale}, it is now possible to tune the phase of W films on the nanoscale. \np

In this work, the local patterning of the W crystalline phase is implemented by DWLA to obtain precise control over the density of $\alpha$-W grains in $\beta$-W thin films, as revealed by transmission electron microscopy (TEM). With electrical measurements, a decrease of the film resistivity from \\230~$\mu\Omega$~cm to 30~$\mu\Omega$~cm is detected, along with a decrease in the spin-orbit torque efficiency of W from 0.45 to $<0.1$ as expected with this crystalline phase transformation. Field-free spin-orbit torque switching in W/CoFeB bilayers is then demonstrated by introducing a gradient in the W phase. These patterned lateral phase gradients in W provide a tunable platform to understand the physical mechanisms of field-free switching in polycrystalline systems with broken structural symmetry. 

\section{Tracking structural transformation of W with laser annealing}
\begin{figure*}[htb!]
    \centering
    \includegraphics[width=0.9\textwidth]{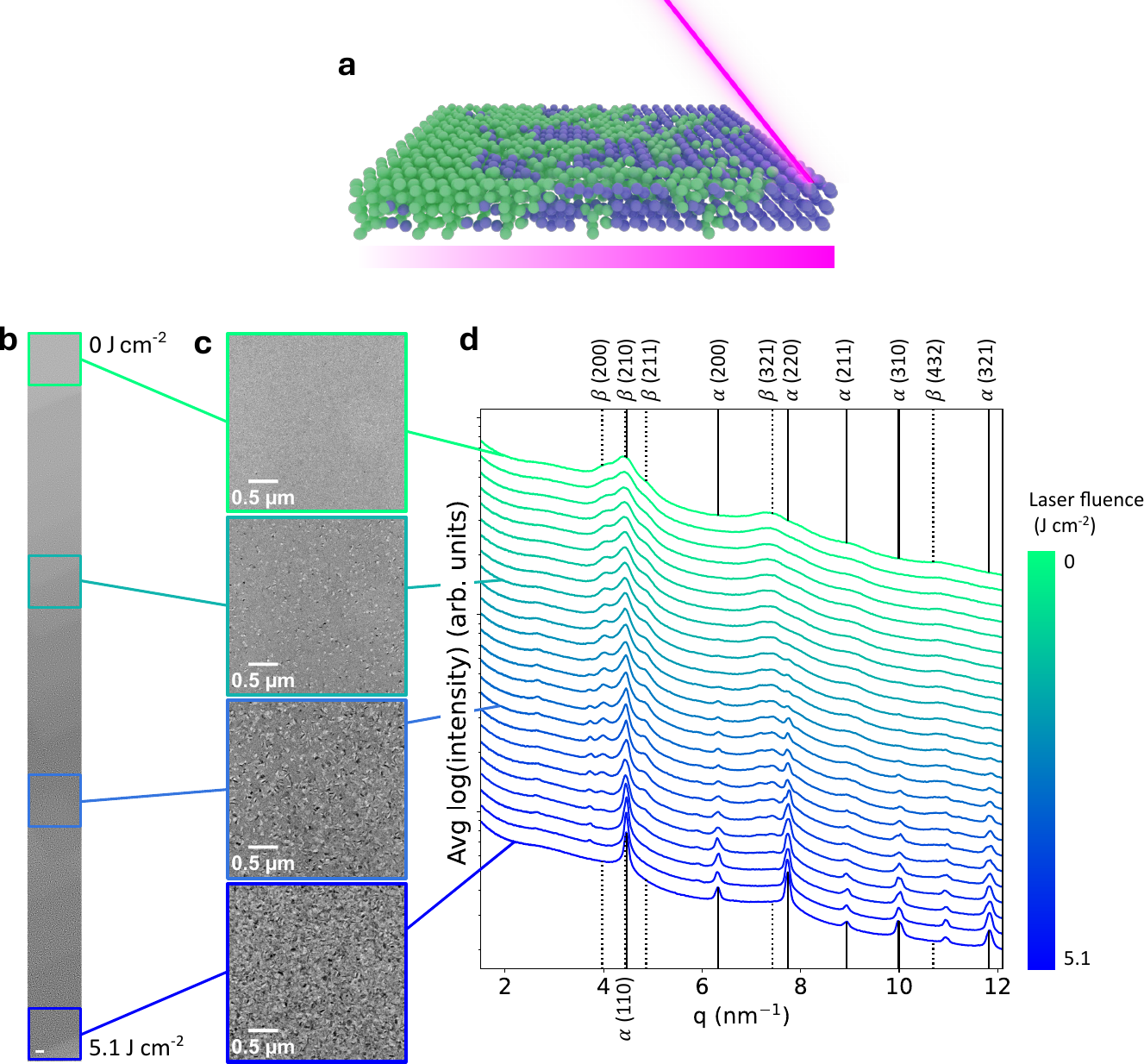}
    \caption{Structural transformation of W thin films. a) Schematic of DWLA of W. As the laser rasters over the W surface, the deposited thermal energy leads to a change in the crystalline phase. The laser fluence is increased linearly from left to right, indicated by the increasing intensity of pink in the bar below the schematic. The green spheres indicate W atoms in the $\beta$-W phase, while the blue spheres indicate W atoms in the $\alpha$-W phase. \hspace{10em} b) A series of stitched TEM images taken at 8000$\times$ magnification along the linear gradient in laser fluence for a 15~nm-thick W film on a silicon nitride membrane (scale bar=0.5~\um). c) Selected TEM images at 8000$\times$ magnification along the gradient. The uppermost panel is at the lowest fluence end of the gradient and appears visually uniform. For the images taken in regions with increasing laser fluence, there are an increasing number of large black and white grains. d) Selected-area electron diffraction performed in 2~\um\ steps along the fluence gradient, with the intensity shown as a function of scattering vector $q$. As the fluence is increased from 0 \jcm\ to 5.1 \jcm, diffraction peaks corresponding to $\beta$-W smoothly transition to diffraction peaks that correspond to $\alpha$-W. }
    \label{fig:fig1tem}
\end{figure*}

Metastable $\beta$-W films are grown by sputtering. The $\beta$-W phase of as-grown films is confirmed with x-ray diffraction, where only diffraction peaks corresponding to the $\beta$-W phase are observed (see Supporting Information Figure~S1). The W films are then locally modified using direct-write laser annealing \cite{riddiford2024grayscale} with laser fluences ranging from 0 \jcm\ to 31 \jcm. The maximum laser fluence used depends on the thermal conductivity of the substrate; for silicon nitride membrane substrates, the W films were annealed at a maximum laser fluence of 5.1 \jcm, while for W films on thermally oxidized Si substrates, the maximum laser fluence was 31 \jcm. In order to examine the microstructural changes in W that result from the local laser treatment, several regions of a 15~nm-thick W film grown on a silicon nitride membrane are annealed (see schematic in Figure~\ref{fig:fig1tem}a). Then, TEM is employed to image the annealed region in the film plane. To track the evolution of the W microstructure, a 50~\um $\times$50~\um\ region in a W film is annealed with a linear gradient in laser fluence. In Figure~\ref{fig:fig1tem}b, TEM images taken along the entire 50~\um\ fluence gradient at 8000$\times$ magnification are stitched together \cite{preibisch2009globally} and shown as a single image. Larger images of four regions along the gradient are shown in Figure~\ref{fig:fig1tem}c. In the as-grown region (uppermost panel), there are only a few sparse large grains visible at this magnification, which appear in black and white due to their diffraction of the electron beam away from or towards the detector, while the gray uniform region is made up of very small grains. With laser annealing, the grain size and density increases as the laser fluence is increased (middle panels). Finally, at the maximum laser fluence, only large W grains are observed, and there are no gray regions remaining (lower panel). This trend is similar to behavior observed in W for in-situ TEM annealing studies \cite{donaldson2018solute}. From TEM images taken at 120,000$\times$ magnification, the size of the small grains are estimated to be 5-20~nm, while the large grains are 80-120~nm in size (seen in Supporting Information Figure~S2a,b).  \np

To gain crystallographic information about the two distinct phases in W, selected-area electron diffraction (SAED) using a 1.2~\um\ aperture was carried out along the gradient in 2~\um\ steps. Rings of intensity in SAED are observed (see Supporting Information Figure~S2c-f), typical of polycrystalline films. Therefore, the radial average of the SAED signal is taken and plotted as a function of the scattering vector $q$ in Figure~\ref{fig:fig1tem}d for each SAED image taken along the gradient. By comparing electron diffraction peaks to the positions of diffraction peaks calculated using Bragg's law for the $\beta$-W and $\alpha$-W crystal structures, it can be seen that the broad diffraction peaks for the as-grown region of the W film correspond only to $\beta$-W peaks, indicative of a uniform $\beta$-W film with very small grains, while the region annealed at the highest laser fluence of 5.1 \jcm\ has only $\alpha$-W peaks, with the narrow peak width typical of larger crystal grains. Between these two fluences, a smooth transition from diffraction data with peaks corresponding only to $\beta$-W to diffraction data with peaks from both phases can be seen, thus indicating a mixed phase in this fluence window. From this, it can be concluded that the gray, small-grain region seen in the TEM images of Figure~\ref{fig:fig1tem}b,c corresponds to $\beta$-W, while the large grains are composed of $\alpha$-W. This thermally-driven phase transformation is notable because, instead of a deformation of the $\beta$-W lattice into an $\alpha$-W lattice, an $\alpha$-W lattice nucleates and expands \cite{barmak2017transformation}, leading to a bimodal distribution of grains in the mixed-phase films \cite{kim2018microstructural}. Thus, a \um-scale gradient will create many structurally asymmetric interfaces on the nanoscale, with 80-120~nm-sized $\alpha$-W grains interfaced with 5-20~nm-sized $\beta$-W grains. 

\section{Electronic and spintronic properties}
\begin{figure}[ht!]
    \centering
    \includegraphics[width=0.68\linewidth]{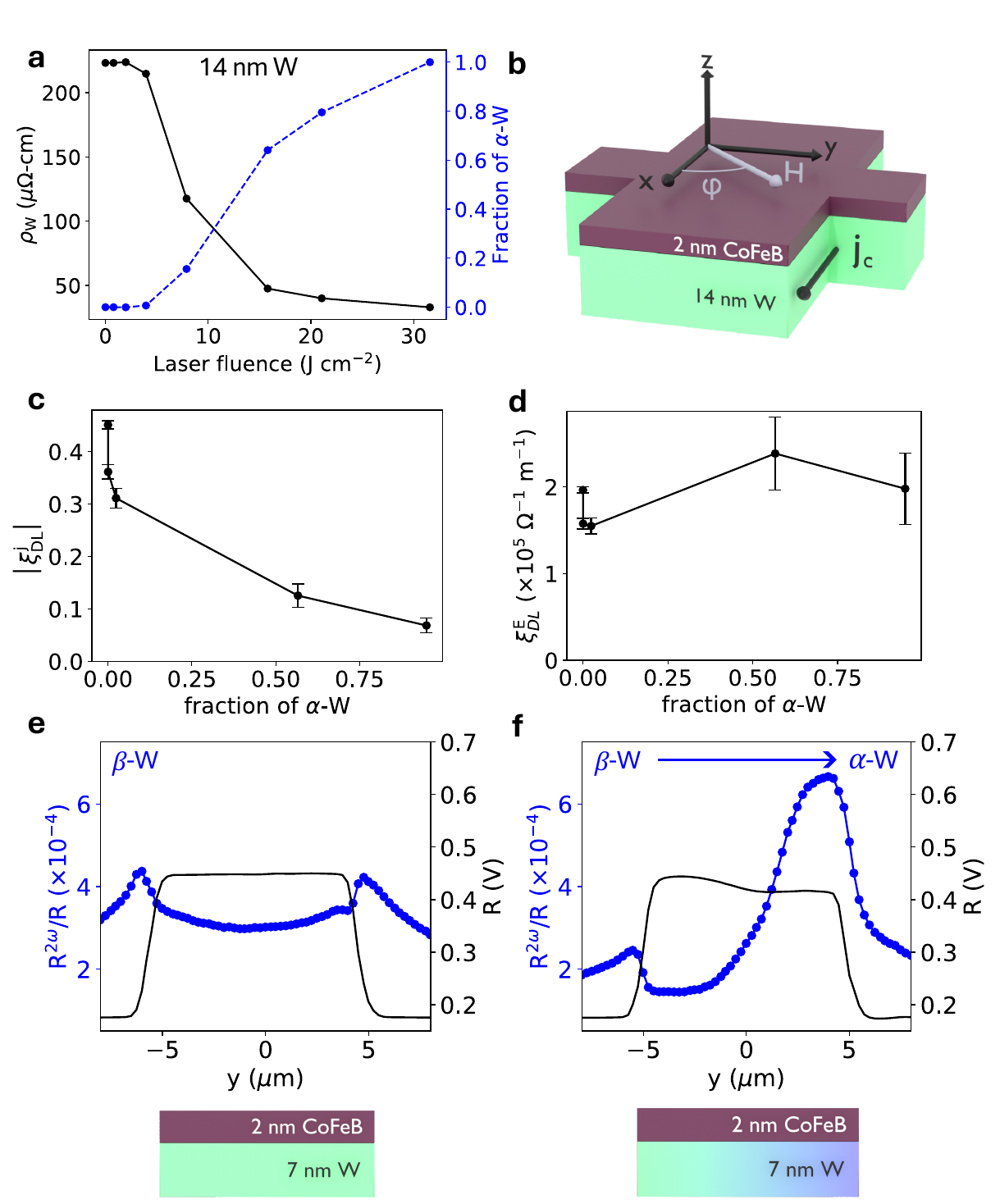}
    \caption{Transformation of electronic and spintronic properties of W on changing the phase with DWLA. \hspace{5em} a) Resistivity $\rho_\text{W}$ of 14~nm-thick W films as a function of laser fluence. Using a parallel resistor model, the fraction of $\alpha$-W as a function of laser fluence is calculated from these resistivity values. b) Measurement geometry for second harmonic Hall measurements. An ac charge current $j_c$ flows along the \textit{x}-direction and the sample is rotated in an external magnetic field $H$. At each angle $\varphi$ between $H$ and \textit{x}, the second harmonic of the resistance along \textit{y} is recorded. c) The magnitude of the damping-like torque efficiency per current $\xi_\text{DL}^j$ as a function of the fraction of $\alpha$-W for 14~nm-thick W. Error bars are given by the standard deviation calculated from the fitting procedure used to determine  $\xi_\text{DL}^j$ (see Methods). d) The damping-like torque efficiency per electric field $\xi_\text{DL}^\text{E}$ as a function of the fraction of $\alpha$-W. Error bars are given by the propagated standard deviation from the fitting procedure used to determine  $\xi_\text{DL}^j$. e-f) Thermoreflectance (R$^{2\omega}$/R) and reflectivity (R) as a function of position across a 10~\um-wide 7~nm W/2~nm CoFeB device. The thermoreflectance (shown in blue), is proportional to the current-induced heating across the device. e) Thermoreflectance of a $\beta$-W device (depicted by the schematic below) is relatively constant across the device . f) For a sample with a gradient from 0\% to 100\% $\alpha$-W (depicted by the color gradient from green to blue in the schematic below), the thermoreflectance is highly non-uniform.}
    \label{fig:electronic}
\end{figure}

The microstructural transformation from $\beta$-W to $\alpha$-W observed in TEM is accompanied by dramatic changes in the electronic and spintronic properties of the films. For as-grown films, the average film resistivity $\rho_W$ is $233$ $\mu\Omega$-cm, comparable to literature values for $\beta$-W \cite{kim2018microstructural}. After annealing above a threshold fluence of 3~\jcm, $\rho_W$ drops significantly (as seen in black in Figure~\ref{fig:electronic}a for a 14~nm-thick W film grown on a thermally oxidized Si substrate). At fluences higher than 20~\jcm, $\rho_W$ stabilizes at 30~$\mu\Omega$-cm, comparable to values for $\alpha$-W films \cite{kim2018microstructural}. The fraction of $\alpha$-W as a function of laser fluence, seen in blue in Figure~\ref{fig:electronic}a, is calculated using the measured resistivity values in black and a parallel resistor model with $\alpha$-W and $\beta$-W. This decrease in $\rho_W$ as the laser fluence is increased eliminates the possibility that there is significant oxidation of the tungsten due to DWLA \cite{riddiford2024grayscale}, since tungsten oxides have a higher resistivity than $\beta$-W \cite{al2023temperature}. \np

The stability of mixed-phase W created with DWLA was investigated by measuring the resistivity of a W Hall bar with a gradient of phase over the span of several months. In contrast to unstable mixed-phase films grown by sputtering, which transformed to $\alpha$-W within days \cite{rossnagel2002phase}, the resistivity of the annealed regions did not exhibit large changes in resistivity over time. For example, a 4\% increase in the resistivity of a W device with a gradient from 0\% to 50\% $\alpha$-W over 10~\um\ was measured over five months (see Supporting Information Figure~S3). Since a transformation of $\beta$-W to $\alpha$-W would result in lower resistivity, this increase indicates there is minor surface oxidation rather than transformation to $\alpha$-W. This means that devices with phase-controlled W regions created with DWLA are stable for long-term use. \np 

The spin-orbit torque efficiency of W also changes significantly depending on the crystalline phase. To verify this, second harmonic Hall resistance measurements of spin-orbit torques were performed on W devices annealed homogeneously with a single laser fluence using CoFeB/MgO/Ta deposited on 14~nm-thick W as the magnetic layer with in-plane magnetization. In this technique, with the sample geometry illustrated in Figure~\ref{fig:electronic}b, an ac current in the bilayer produces current-induced torques leading to oscillations of the CoFeB magnetization. These in turn lead to an oscillating Hall resistance that can be detected by measuring the second harmonic Hall resistance (R$_{xy}^{2\omega}$). By fitting the angular, magnetic field, and current dependence of R$_{xy}^{2\omega}$, the damping-like torque efficiency per current $\xi_\text{DL}^j$ and per electric field $\xi_\text{DL}^\text{E}$ of W can be extracted \cite{garello2013symmetry, avci2014interplay} (see Supporting Information Figure~S4 for example data). $\xi_\text{DL}^j$ is a figure of merit describing how efficiently charge currents are converted to spin currents in a material and is related to the commonly cited spin Hall angle $\theta_\text{SH}$ by $\xi_\text{DL}^j=T_\text{int}\theta_\text{SH}$, where $T_\text{int}$ is the spin transparency at the heavy metal/magnet interface \cite{zhu2019maximizing}. \np

$\xi_\text{DL}^j$ as a function of $\alpha$-W fraction is shown in Figure~\ref{fig:electronic}c. $\beta$-W has $\xi_{DL}^j=-0.45\pm0.01$, which is comparable to measured values of $\theta_{SH}$ \cite{pai2012spin, vudya2021optimization, hao2015beta, liu2015correlation}. However, small fractions of $\alpha$-W lead to a steep decrease of $\xi_\text{DL}^j$, which may be exacerbated by a decreased interface quality between W and CoFeB after annealing at low laser fluences. $\xi_\text{DL}^j$ continues to decrease with increasing fraction of $\alpha$-W, albeit more gradually, and at 95\% $\alpha$-W, $\xi_{DL}^j=-0.07\pm0.01$. $\xi_\text{DL}^\text{E}$ (given by $\xi_\text{DL}^j/\rho_\text{W}$) as a function of $\alpha$-W fraction (seen in Figure~\ref{fig:electronic}d) provides a scaled spin-orbit torque efficiency accounting for non-uniform current flow through a W wire with a gradient in phase. Here, devices with larger fractions of $\alpha$-W have a relatively high $\xi_\text{DL}^\text{E}$ due to the significantly lower resistivity, although the large measurement error obscures a clear trend in $\xi_\text{DL}^\text{E}$ as a function of $\alpha$-W fraction. \np

To directly visualize the impact of adding a gradient in the W phase on the electronic properties of a device, the thermoreflectance of 10~\um-wide 7~nm W/2~nm CoFeB devices was detected with current-modulated, spatially-resolved optical reflectivity measurements (see Methods). The reflectivity of the devices was measured as a laser scanned along the \textit{y}-direction transverse to an ac current applied to the devices along \textit{x}. The magnitude of the time-averaged reflectivity $R$ is determined by the microstructure of the film. The thermoreflectance ($R^{2\omega}/R$) is defined as the second harmonic reflectivity change $R^{2\omega}$ normalized by $R$. Applying an ac current results in harmonic Joule heating, which in turn leads to a harmonic variation of the refractive index. Hence, the thermoreflectance is a measure of the Joule heating-induced change in reflectivity, giving a qualitative insight into the current density~\cite{mohan2025time}. A device with uniform, as-grown $\beta$-W, shown in Figure~\ref{fig:electronic}e, is compared to a device with a gradient in phase going from 0\% to 100\% $\alpha$-W over 10~\um, shown in Figure~\ref{fig:electronic}f. The reflectivity (black line in Figure~\ref{fig:electronic}e-f) gives an indication of the edges of the device, and regions with more $\alpha$-W have a slightly lower reflectivity than $\beta$-W, seen by the black reflectivity curve for the gradient device in Figure~\ref{fig:electronic}f. The thermoreflectance of the $\beta$-W device is relatively constant (Figure~\ref{fig:electronic}e), reflecting the uniform current flow across the as-grown W. In contrast, the effect of strongly spatially-varying resistivity is exemplified by the thermoreflectance of the gradient device, depicted in blue in Figure~\ref{fig:electronic}f.  Due to the low resistivity of $\alpha$-W, the vast majority of the current passes through the far right region of the gradient. These results indicate that landscapes with smooth gradients in the spin-orbit torques and current density of W can be created. \np

\section{Spin-orbit torque switching}
\begin{figure}[ht!]
    \centering
    \includegraphics[width=0.65\linewidth]{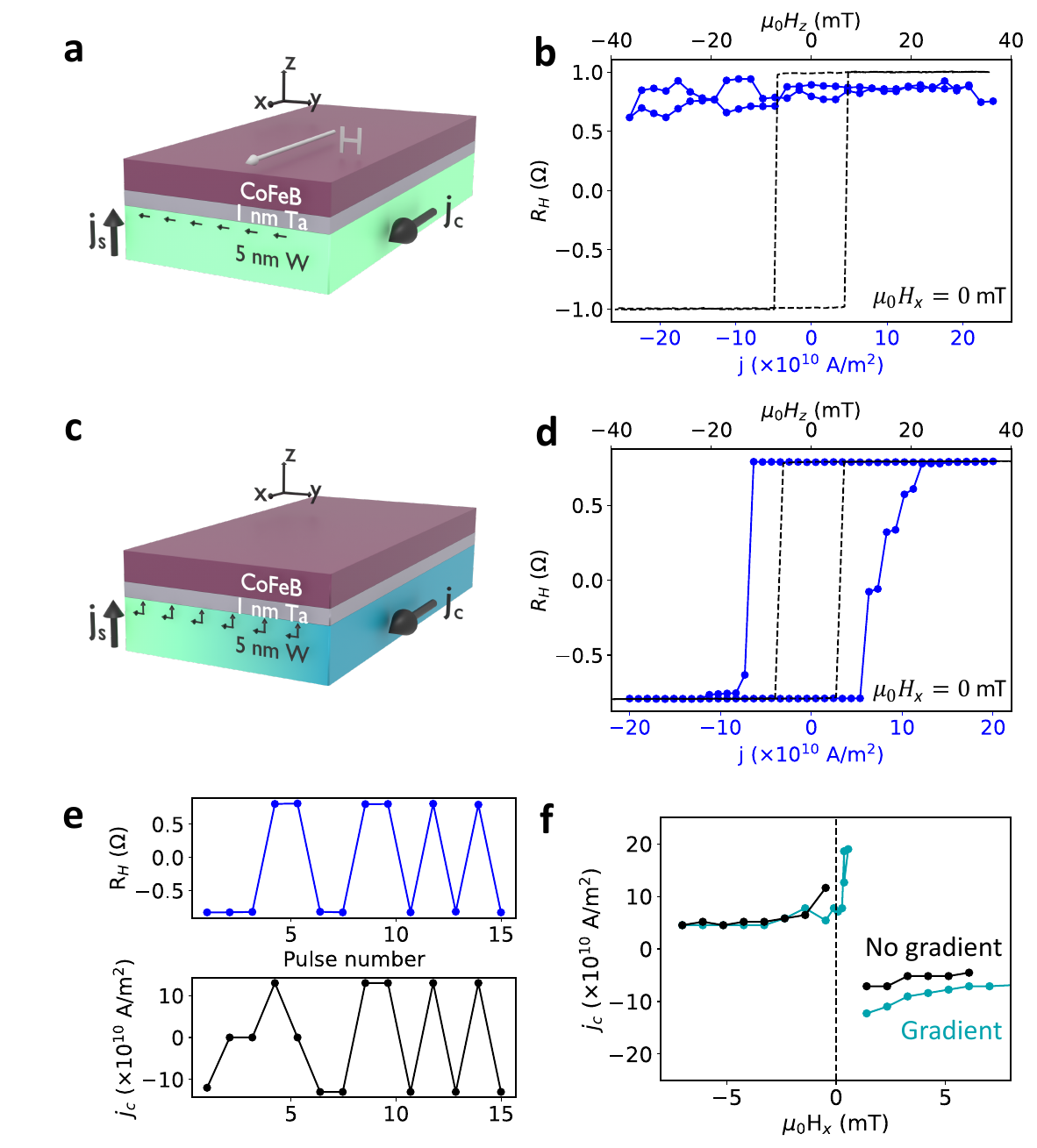}
    \caption{(a) Schematic of spin-orbit torque switching in $\beta$-W/Ta/CoFeB. A current $j_c$ is applied along the $x$-direction. The spin Hall effect in W generates a spin current $j_s$ with $y$-polarized spins due to symmetry about the $xz$ plane. (b) $R_\text{H}$ of $\beta$-W/Ta/CoFeB as a function of $\mu_0 H_z$ (black dashed line) and the change in $R_\text{H}$  with $j$ with no applied magnetic field (in blue). (c) Schematic of spin-orbit torque switching with a gradient in W phase from $\beta$-W (in green) to mixed $\beta$-W/$\alpha$-W phase (in blue). (d) $R_\text{H}$ as a function of $\mu_0 H_z$ (black dashed line) and change in $R_\text{H}$ as a function of $j$ (in blue) without any applied in-plane field for a device with a gradient in W phase.\hspace{5em} (e) Reliability of field-free current-induced switching of CoFeB on W with a phase gradient. In the upper panel, $R_\text{H}$ of the device is recorded after a series of current pulses with the magnitude shown in the bottom panel. (f) $\mu_0 H_x$ dependence of $j_c$ for devices with (in turquoise) and without (in black) W phase gradients. }
    \label{fig:switching}
\end{figure}

To perform SOT switching, elliptical dots with minor (major) axis of 5.7~\um\ (10~\um) of 1~Ta/\\ 0.9~CoFeB/1~MgO/3~Ta (listed from the lowest layer on the substrate to the top layer, with all thicknesses in nm) are deposited on laser-annealed W devices. The 1~nm-thick Ta seed layer is included to stabilize perpendicular magnetic anisotropy in the CoFeB layer without the need for any post-growth annealing. 50~\textmu s current pulses are applied to the devices for spin-orbit torque switching experiments, and the Hall resistance $R_\text{H}$ is recorded after each current pulse to detect the magnetization orientation (see Methods). $R_\text{H}$ is also measured as a function of applied out-of-plane magnetic field $\mu_0 H_z$ in order to detect the $R_\text{H}$ values corresponding to uniform up/down magnetization. For devices uniformly annealed at a constant laser fluence, the critical switching current density $j_c$, defined as the average of the two current densities where $R_\text{H}$ reverses sign from negative to positive and from positive to negative, at a given in-plane magnetic field follows the trend observed in second harmonic Hall measurements of lower spin-orbit torque efficiency for devices with a higher fraction of $\alpha$-W (see Figure~\ref{fig:electronic}c), where the switching current increases significantly for devices annealed at higher laser fluence (see Supporting Information Figure~S5a). For these uniformly annealed devices and devices with an unannealed $\beta$-W film, complete magnetization reversal is only achieved for magnetic fields greater than 1 mT applied along the current direction due to the symmetry of the heterostructure (see Figure~\ref{fig:switching}a), consistent with previous reports \cite{miron2011perpendicular}. Without an applied in-plane field, no switching of the $\beta$-W device is observed, as seen in Figure~\ref{fig:switching}b with $R_\text{H}$ as a function of $\mu_0 H_z$ shown as a black dashed line, and $R_\text{H}$ as a function of the applied current density $j$ shown in blue.\np

Devices with a gradient in the W phase created transverse to the current direction (depicted by the schematic in Figure~\ref{fig:switching}c) exhibit different behavior than the uniform devices. For a device with a gradient in phase going from 0\% to 20\% $\alpha$-W across the 10~\um\ width, the magnetization fully switches at $j=1.4\times10^{11}$ A/m$^2$ without an applied magnetic field. Here, the current density in Figure~\ref{fig:switching}d-f is defined as the total applied current divided by the cross-sectional area. In Figure~\ref{fig:switching}d, $R_\text{H}$ as a function of $j$ with no applied magnetic field (in blue) is compared to $R_\text{H}$ as a function of $\mu_0 H_z$ (black dashed line), and it is seen that $R_\text{H}$ switches between uniform up (down) magnetization states with positive (negative) current pulses. The reproducibility of this field-free switching is tested by applying a pattern of current pulses, seen in the lower panel of Figure~\ref{fig:switching}e, and measuring $R_\text{H}$, seen in the upper panel of Figure~\ref{fig:switching}e. It is observed that $R_\text{H}$ reliably switches to the maximum (minimum) values measured in Figure~\ref{fig:switching}d with positive (negative) current pulses, indicating reproducible and complete magnetization reversal. $j_c$ as a function of applied in-plane magnetic field $\mu_0 H_x$ is compared for a gradient device and as-grown device with no gradient in Figure~\ref{fig:switching}f. $j_c$ at $\mu_0 H_x = 0$~ mT (indicated by the vertical black dashed line in Figure~\ref{fig:switching}f) for the gradient device is positive, matching the polarity of $j_c$ for the uniform device with $-\mu_0 H_x$ applied. This suggests that the gradient creates an effective $-\mu_0 H_x$ that aids field-free switching. \np

\begin{figure}[ht!]
    \centering
    \includegraphics[width=0.68\linewidth]{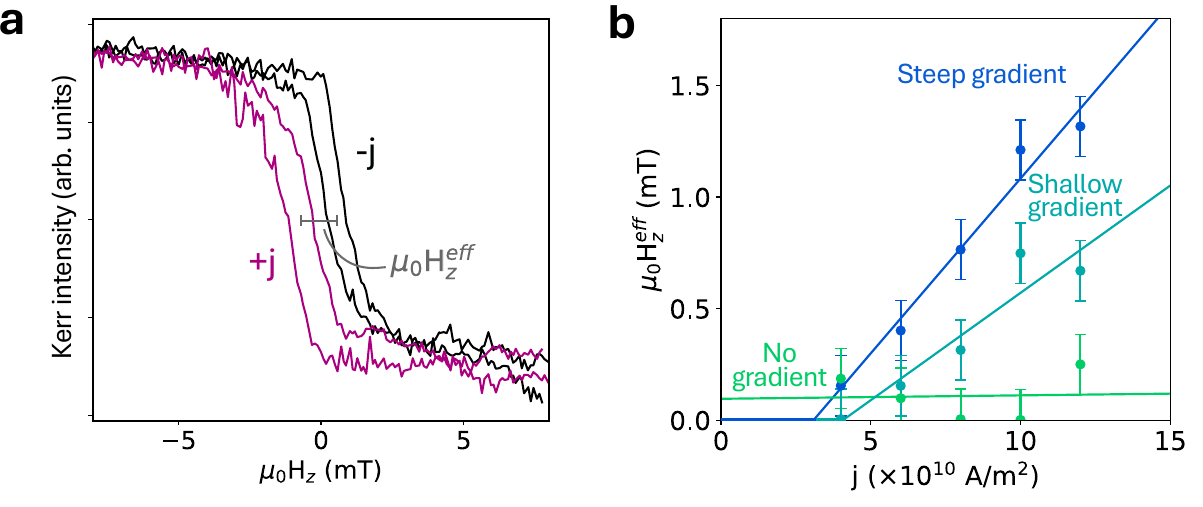}
    \caption{a) Example of hysteresis loop shift measurements performed at a current density of $1\times10^{11}$ A/m$^2$ for a 0\%-20\% $\alpha$-W gradient across 10~\um. The purple (black) hysteresis loop is for positive (negative) current direction. The difference between the center of the loops for positive and negative current is defined as $\mu_0 H_z^\text{eff}$. b) The current dependence of $\mu_0 H_z^\text{eff}$ for a steep gradient (0\%-20\% $\alpha$-W) and a shallow gradient (0\%-10\% $\alpha$-W) in $\beta$-W over 10~\um. A device without a gradient exhibits no loop shift. Error bars are given by the magnetic field step size used to collect hysteresis loops. }
    \label{fig:moke}
\end{figure}

To verify the presence of unconventional torques, which would produce this field-free switching, measurements of hysteresis loop shifts are performed for three W/Ta/CoFeB devices \cite{baek2018spin}: one with a gradient in the $\beta$-W phase with $\alpha$-W going from 0\% to 20\% over the 10~\um\ device width (the steep gradient), one with a gradient from 0\% to 10\% $\alpha$-W (the shallow gradient), and one with no gradient. Here, a dc current is passed through the device and the polar MOKE response as a function of $\mu_0 H_z$ is measured in a Kerr microscope, with hysteresis loops at applied dc currents of $\pm1\times10^{11}$ A/m$^2$ shown in Figure~\ref{fig:moke}a. The shift in magnetic field of the hysteresis loop between positive and negative dc current, termed $\mu_0 H_z^\text{eff}$, without an applied magnetic field along \textit{x} is characteristic of unconventional torques in the system. For the device with no gradient, there is no loop shift outside the error of the measurement  (the green data in Figure~\ref{fig:moke}b). Meanwhile, both gradient devices (the turquoise and blue data in Figure~\ref{fig:moke}b) have a measurable $\mu_0 H_z^\text{eff}$ that increases with increasing current density. It is also observed that, for both gradient devices, there is a threshold current below which there is no measurable hysteresis loop shift, similar to a film with a gradient in Ta phase \cite{zeng2025deterministic} and films where low crystal symmetry generates z-polarized spins \cite{hu2022efficient, cao2023anomalous}. In addition, the steep gradient has both a lower current threshold for the onset of a nonzero $\mu_0 H_z^\text{eff}$ and a stronger dependence of $\mu_0 H_z^\text{eff}$ on the current density. This indicates that the steep gradient has stronger unconventional torques that overcome the damping of the CoFeB at a lower current density than the shallow gradient. \np

To determine the mechanism of unconventional torque generation, two possibilities are considered. First, $z$-polarized spins can be generated in materials with broken spatial symmetry, leading to unconventional torques that can deterministically switch a neighboring magnet. Second, in the case of W films with a phase gradient, there are several asymmetries in properties over the device which may contribute:
\begin{enumerate}[label=(\roman*)]
    \item There is inhomogeneous current flow through the W due to the differing resistivity of the $\beta$-W and mixed-phase W. The patterned gradients designed for field-free switching have (at most) 0\% to 20\% $\alpha$-W fractions, so the maximum resistivity change would be 230~$\mu\Omega$-cm to 105~$\mu\Omega$-cm from one edge to the other (see Figure~\ref{fig:electronic}a, for a fluence of $\sim 10$ \jcm).
    \item The anisotropy strength of the CoFeB depends on the W phase and thus will differ across a gradient device. In order to quantify this asymmetry, planar Hall effect measurements of CoFeB with the W annealed at constant fluences are used to extract the effective magnetic\\ anisotropy as a function of laser fluence. As the fraction of $\alpha$-W increases, the effective magnetic anisotropy strength decreases by at most 25\% (see Supporting Information Figure~S5a,b).
    \item There are thermal gradients produced by the non-uniform current flow and differing thermal conductivities between $\alpha$-W and $\beta$-W.
\end{enumerate}
These non-uniform physical properties could combine with the spin-orbit torque gradient to switch the magnetization without an applied field.  To examine the relative likelihood of each of these possibilities, a combination of COMSOL Multi\-physics$^{\circledR}$ software \cite{comsol} and MuMax3 \cite{vansteenkiste2014design} are used to simulate W/CoFeB properties. For these simulations, a current density of $3\times10^{11}$ A/m$^2$ through the W layer and a gradient from 0\% to 100\% $\alpha$-W over 10~\um\ were employed to test if the combined current density gradient, anisotropy gradient, and inhomogeneous Oersted field are sufficient for field-free switching. This simulated gradient is steeper than the gradients used in the experiments for field-free switching in order to observe the largest possible changes in current-induced asymmetries across the device. In these simulations, no deterministic magnetization reversal is observed (see Supporting Information section S5 and Supporting Information Figures S6 and S7 for modeling of current density, non-uniform Oersted field, and micromagnetic simulations). While there is a contribution from thermal gradients, it is not expected to be sufficient for field-free switching based on published works with current density gradients, where an estimated 40 K temperature gradient did not play a dominant role in magnetization reversal \cite{kateel2023field}.  \np

While anisotropy and current density gradients are not adequate to produce field-free switching in micromagnetic simulations, deterministic field-free switching is demonstrated in simulations by including $z$-polarized spins (see Supporting Figure~S7). This raises the question of what the underlying physical origin is for $z$-polarized spin generation in polycrystalline films where each of the two crystalline phases is high-symmetry. Broken electronic symmetries have recently been proposed as a way to generate $z$-polarized spins \cite{liu2024efficient}, and the resistivity difference in the W phases may create this electronic asymmetry. Another possibility is an asymmetry resulting from the change in microstructure on annealing. As seen in the TEM images, while the $\alpha$-W and $\beta$-W unit cell are each high-symmetry, there are many grain boundaries with an $\alpha|\beta$ interface (see Figure~\ref{fig:fig1tem}), creating lateral interfaces between different crystal structures and spin Hall conductivities \cite{xue2020staggered}, which may provide the necessary asymmetry for $z$-polarized spin generation.  \np

\section{Conclusions}
Local patterning of the crystalline phase of W with direct-write laser annealing provides a means to tune spin-orbit torque efficiency and current-induced magnetization switching. The density and size of $\alpha$-W grains in a $\beta$-W film can be adjusted on the nanoscale and, as a result, the film resistivity can be modified by almost an order of magnitude, and the strength of spin-orbit torques can be tuned from 0.45 to $\sim0.07$. Gradients in these properties are leveraged to switch the magnetization of a neighboring CoFeB layer by current without an applied magnetic field. The large resistivity difference between phases can be employed to precisely control the current magnitude and direction at all points in a wire. By combining the field-free switching shown here with non-uniform current directions, ultrahigh efficiency of spin-orbit torque switching is likely \cite{vlasov2022optimal}, providing a new route for energy-efficient spintronic devices. Furthermore, given the broad applications of W films across various industries, the ability to locally tune the crystalline phase of a film with accompanying changes in resistivity, thermal conductivity, superconductivity, and optical properties holds great promise for applications beyond spintronics. 

\section{Experimental Section}
\subsection{Film growth}
5-15~nm-thick W films are grown by magnetron sputtering on Si substrates with 100~nm-thick SiOx coating. For TEM measurements, the films are instead grown on a 200~nm-thick silicon nitride membrane. To ensure $\beta$-phase growth in films thicker than 6~nm, the chamber was exposed to 1~sccm N$_2$ flow for 5~minutes before growth to increase the chamber base pressure, as high sputtering base pressure is known to assist stabilization of the $\beta$-W phase \cite{lu2024enlarged, liu2016topologically}. Because the laser fluence required to change the W phase is quite large and would likely damage the magnetic properties of any magnet grown underneath W, the W is always grown on the substrate and annealed first, then placed back into the chamber for growth of any additional layers. 

\subsection{Transmission electron microscopy}
15~nm-thick W films were sputtered on a 200~nm-thick silicon nitride membrane for TEM measurements. After growth, the films were laser annealed. Because the thermal conductivity of the membrane is significantly lower than that of the 500~\um-thick thermally-oxidized Si substrate, the laser fluences used on the membrane were also much lower (5.1 \jcm\ compared to the 31 \jcm\ used on the thermally-oxidized Si substrate) to reproduce the same transformations from $\beta$-W to $\alpha$-W. The equivalence in terms of the transformations obtained by these laser fluences was verified by comparing x-ray diffraction on annealed W films grown on the thermally-oxidized Si substrates to electron diffraction of annealed regions of W films grown on silicon nitride membranes. 

\subsection{Electrical measurements}
For resistivity measurements, 10~\um-wide Hall bars with a 15~\um\ distance between the longitudinal voltage contacts are fabricated from W films on thermally oxidized Si substrates using UV photolithography in combination with liftoff. The resistivity of each Hall bar device is measured by passing through a 0.1 mA current and measuring the voltage across two longitudinal voltage leads. Then, different Hall bars are annealed uniformly with DWLA at different laser fluences. The resistivity of each device is then re-measured such that the resistivity can be directly compared before and after annealing. \np

For second harmonic Hall measurements, 10~\um-wide W Hall bars with a 10~\um\ distance between the longitudinal voltage contacts are uniformly annealed at different laser fluences. Then, a window in photoresist above the Hall bar is defined with photolithography. The W surface within the window is cleaned with low-power Ar plasma etching. Subsequently, an in-plane magnetized 2~nm CoFeB/1~nm MgO/3~nm Ta film is sputter deposited, and then all unwanted material outside the window is removed with liftoff. This sequence of fabrication steps, with the CoFeB film deposited last, ensures that the magnetic properties are not altered by the laser annealing. For spin-orbit torque switching measurements, the 10~\um$\times$10~\um\ W Hall bars are annealed with a laser fluence gradient transverse to the current direction, producing a phase gradient from pure $\beta$-W to mixed phase W. Subsequently, the W surface is cleaned with Ar plasma etching and an out-of-plane magnetized 1~Ta/0.9~CoFeB/1~MgO/2~Ta film is deposited on the W film. The Ta seed layer is used to ensure that the CoFeB has perpendicular magnetic anisotropy as-grown. The CoFeB elliptical dots with minor (major) axis of 5.7~\um\ (10~\um) are then defined with photolithography and Ar plasma etching. After applying a current pulse, the anomalous Hall resistance is measured with a small test current of 0.1 mA.\np

\subsection{Thermoreflectance measurements}
The reflectivity and thermoreflectance of films was probed with a 520~nm-wavelength laser beam that was focused through an objective lens (100$\times$ magnification, NA$=0.9$), resulting in a spot size of $\sim 0.5$~\textmu m~\cite{stamm2017magneto}. The incident beam was set to be linearly polarized and at normal incidence to the film's surface. The reflected light intensity $R$ was detected using a Thor labs DET36A photodetector. The laser beam was scanned along a line transverse to the current direction.\np

Static (time-averaged) and dynamic responses of the system to a current-induced stimulus are resolved by applying alternating (sine wave) currents and demodulating the signal from the photodetector using a lock-in amplifier (LIA). The harmonic changes of the reflectivity are linked to the temperature dependence of the film's refractive index $n$. The temperature is modified due to Joule heating which scales with the square of applied current density~\cite{mohan2025time}. Therefore, the harmonic change in the reflectivity is completely captured by the second harmonic signal from the photodiode $R^{2\omega}$. Thermoreflectance relates the normalized change in reflectivity $R^{2\omega}/R$ to the change in temperature $\Delta T$ through: $R^{2\omega}/R=c_\mathrm{TR} \Delta T $ , where $c_\mathrm{TR}$ is the material and wavelength dependent thermoreflectance coefficient~\cite{favaloro2015characterization}. 
\np

\medskip

\textbf{Acknowledgements} \par 
Each author contributed to this work as follows: L.J.R. and A.H. conceived the project with input from L.J.H. L.J.R. and A.F. synthesized, laser annealed, and characterized W films and devices. A.F. performed x-ray diffraction and resistivity measurements. E.M. performed TEM measurements, and the TEM data was analyzed by L.J.R. S.D. performed second harmonic Hall resistance measurements, and the data was analyzed by S.D. and L.J.R. E.K., N.K., and T.G. performed reflectivity and thermoreflectance measurements and analyzed the data, with help from P.G. L.J.R. performed spin-orbit torque switching and loop shift measurements and analyzed data with input from A.H. and P.G. L.J.R. wrote the manuscript with help from A.H. and L.J.H. All authors contributed to finalizing the manuscript. L.J.R acknowledges funding from the ETH Zurich Postdoctoral Fellowship Program 22-2 FEL-006. The Laboratory for Magnetism and Interface Physics acknowledges support from SNF Grant No. 200021-236524. We thank the cleanroom staff at the Laboratory for Nano and Quantum Technologies at the Paul Scherrer Institute for technical support. We acknowledge the usage of the instrumentation provided by the Electron Microscopy Facility (EMF) at PSI and we thank the EMF team for their help and support. We also thank Min-gu Kang, Paul No\"{e}l, and Aur\'elien Manchon for helpful discussions. The raw data that supports this study will be available at the Zenodo repository \url{https://doi.org/10.5281/zenodo.17937763}.

\medskip

\bibliographystyle{unsrt}
\bibliography{refs}

\newpage
\setcounter{section}{0}
\setcounter{equation}{0}
\setcounter{table}{0}
\setcounter{figure}{0}
\renewcommand{\thefigure}{S\arabic{figure}}
\renewcommand{\thetable}{S\arabic{table}}
\renewcommand{\thesection}{S\arabic{section}}
\renewcommand{\thesubsection}{S\thesection.\arabic{subsection}}
\renewcommand{\thesubsubsection}{S\thesubsection.\arabic{subsubsection}}

\begin{center}
{\huge \textbf{Supporting Information}} \\
\vspace{0.5em}
{\LARGE Generating unconventional spin-orbit torques with patterned phase gradients in tungsten thin films} \\
\vspace{2em}
{\Large Lauren J. Riddiford\footnote{lauren.riddiford@psi.ch, laura.heyderman@psi.ch, ales.hrabec@psi.ch}, Anne Flechsig, Shilei Ding, Emir Karadza, Niklas Kercher, Tobias~Goldenberger, Elisabeth M\"{u}ller, Pietro Gambardella, Laura~J.~Heyderman$^*$, Ale\v{s} Hrabec$^*$} 
\end{center}
\section{Further structural characterization}
\begin{figure}[h!]
    \centering
    \includegraphics[width=0.8\linewidth]{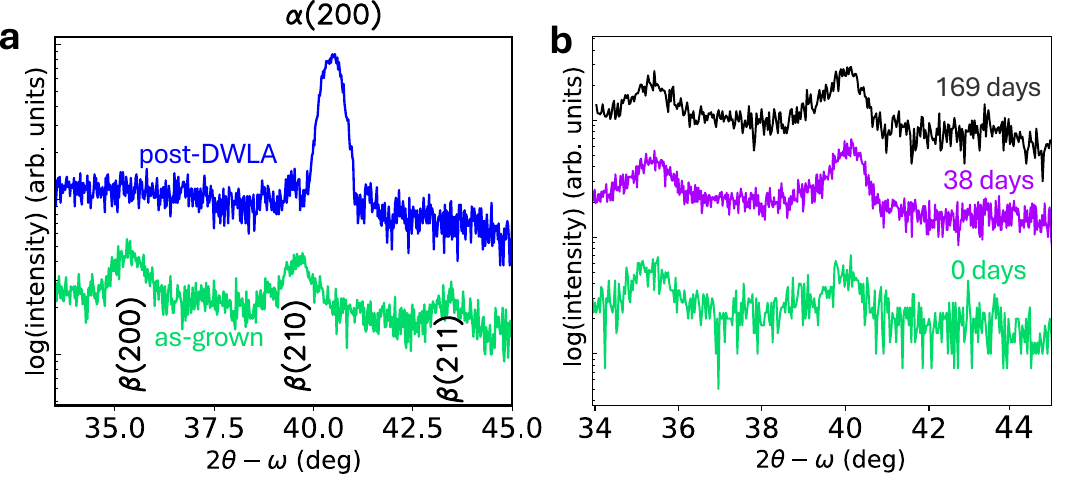}
    \caption{a) X-ray diffraction scans for 14 nm-thick W films grown on thermally oxidized Si substrates before (in green) and after (in blue) laser annealing at $\sim25$~\jcm. In the as-grown film, only film peaks corresponding to $\beta$-W are detected. After laser annealing, only the $\alpha$-W diffraction peak is visible. b) X-ray diffraction scans of an as-grown 14 nm-thick W film recorded at 0, 38, and 169 days after growth. For all scans, only the $\beta$-W film peaks are visible. This confirms that the metastable $\beta$-W films are stable over several months. }
    \label{fig:xrd}
\end{figure}

X-ray diffraction scans of 14 nm-thick W films grown on thermally oxidized Si substrates confirm that the films, as-grown, are composed of only $\beta$-W, as seen in Figure~\ref{fig:xrd}a (green data). After laser annealing at high fluence of $\sim25$~\jcm, the film can be transformed completely to $\alpha$-W, confirmed by the disappearance of $\beta$-W peaks and the appearance of a strong $\alpha$-W peak (blue data). The stability of the as-grown $\beta$-W films was confirmed by measuring a series of x-ray diffraction scans over several months. It is observed that, even 5.5 months after growth, only $\beta$-W diffraction peaks are detected in XRD data, seen in Figure~\ref{fig:xrd}b. 

\begin{figure}[h!]
    \centering
    \includegraphics[width=0.95\linewidth]{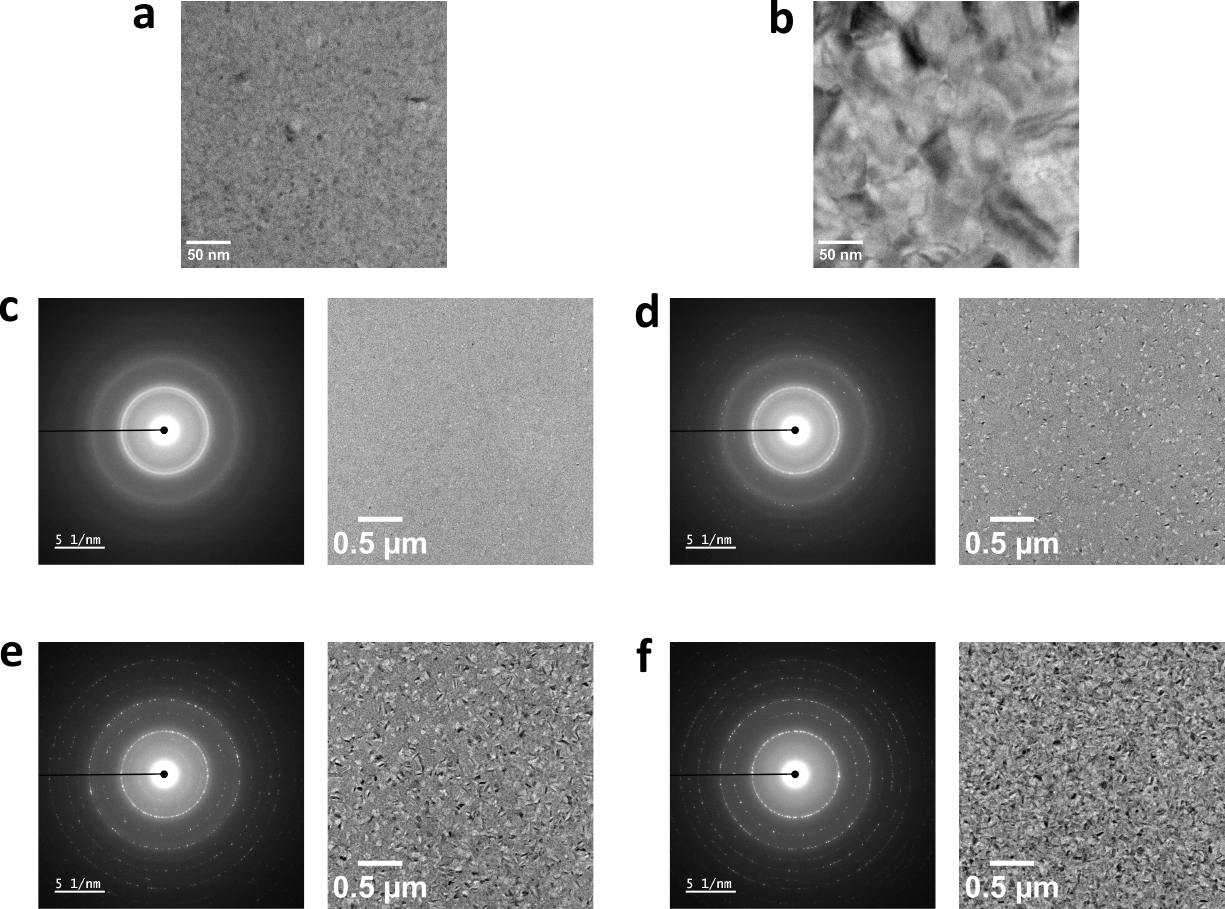}
    \caption{a) TEM image of an as-grown $\beta$-W film. b) TEM image of an annealed, $\alpha$-W region in W film. (c-f) SAED and associated TEM images for four locations along the gradient which correspond to the four images shown in Figure 1d: c) the as-grown film, d-e) mixed-phase films with increasing laser fluence, and f) W film at the maximum laser fluence of 5.1~\jcm, which has entirely transformed to $\alpha$-W. The film used for these TEM images is grown on a silicon nitride membrane.}
    \label{fig:tem2}
\end{figure}

\clearpage
The grain sizes of $\beta$-W and $\alpha$-W are estimated by measuring individual grains in transmission electron microscopy (TEM) images taken at 120,000$\times$ magnification. $\beta$-W grains are quite small, with sizes measuring from $\sim5-20$~nm (Figure~\ref{fig:tem2}a), while $\alpha$-W grains are are $\sim80-120$~nm in size (Figure~\ref{fig:tem2}b). The diffraction intensity as a function of scattering vector shown in Figure 1e was generated by taking the radial average of selected-area electron diffraction (SAED) images taken at different locations along the gradient. Four representative SAED images, with each next to the TEM images of the area where they were recorded, are shown in Figure~\ref{fig:tem2}c-f. The aperture for the SAED images was 1.2 \textmu m. 
\clearpage
\section{Electrical characterization of mixed phase stability}
\begin{figure}[h!]
    \centering
    \includegraphics[width=0.5\linewidth]{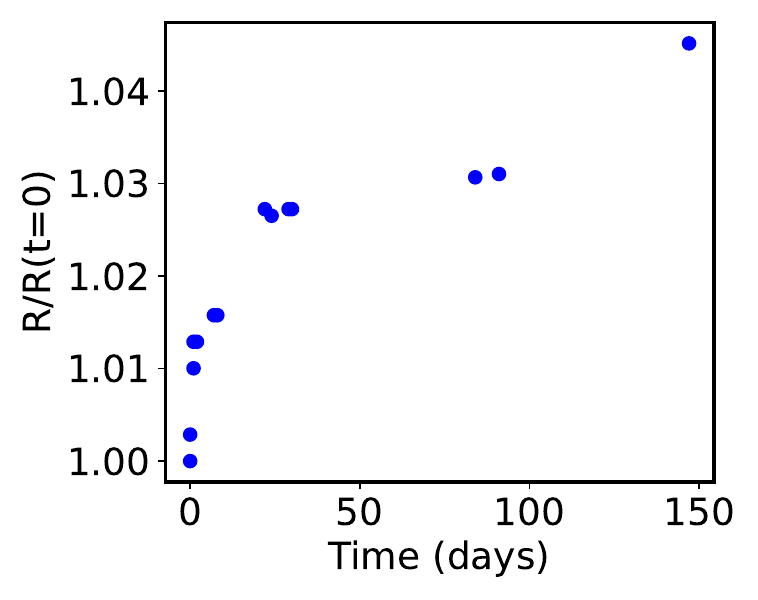}
    \caption{The resistance of a W wire with an annealed gradient is tracked over 150 days. Only a small ($\sim4.5$\%) increase in resistance is measured in this time frame. }
    \label{fig:res_stability}
\end{figure}

The longitudinal resistance $R$ of a 10 \textmu m-wide and 10 \textmu m-long W wire with a lateral gradient in $\beta$-W, going from 0\% to 50\% $\alpha$-W, was tracked over 150 days. After 150 days, the resistance had increased by $\sim4.5$\%, as shown in Figure~\ref{fig:res_stability}. Since a transformation of $\beta$-W to $\alpha$-W would cause a decrease in resistance, this confirms that annealed, mixed-phase W remains stable, and the nucleated grains of the stable $\alpha$-W crystalline phase do not cause the remainder of $\beta$-W to transform to $\alpha$-W. The small increase in resistance is attributed to surface oxidation of W. 

\clearpage

\section{Details of second harmonic Hall measurements}
\begin{figure}[h]
   \centering
    \includegraphics[width=0.85\linewidth]{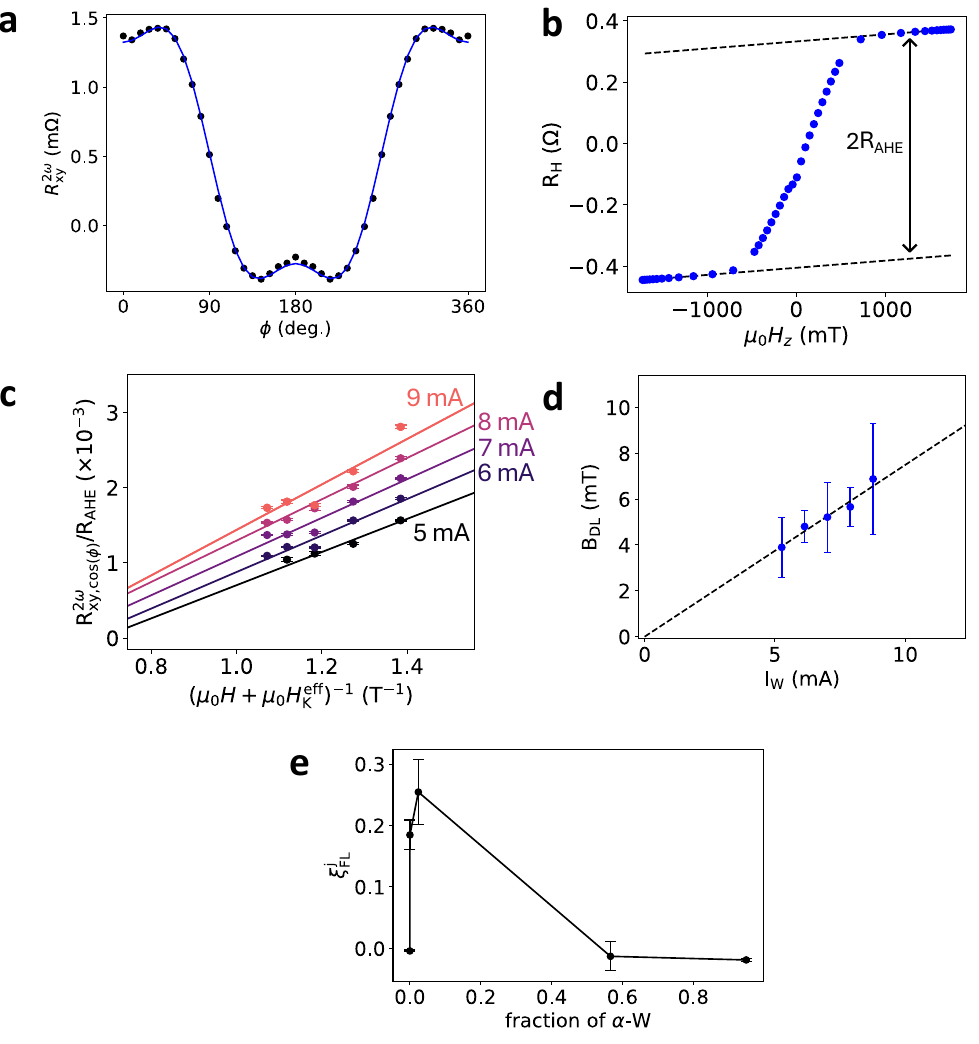}
    \caption{a) $R_\text{xy}^{2\omega}(\varphi)$ for $\beta$-W/CoFeB for an applied magnetic field of 128 mT. b) $R_\text{H}$ of $\beta$-W/CoFeB as a function of $\mu_0H_z$. $R_\text{AHE}$ and $\mu_0 H_\text{K}^\text{eff}$ are extracted from this data. c) The scaled $\cos\varphi$ component of $R_\text{xy}^{2\omega}(\varphi)$ as a function of $(\mu_0 H + \mu_0 H_\text{K}^\text{eff})^{-1}$. Data is plotted for five different currents from 5 mA to 9 mA. d) The change in $B_\text{DL}$ with $I_\text{W}$. The slope of the linear relationship between the two (dashed black line) is proportional to $\xi_\text{DL}^j$. e) An estimate of the magnitude of $\xi_\text{FL}^j$ as a function of $\alpha$-W fraction.}
    \label{fig:2ndharm}
\end{figure}

The spin-orbit torque efficiency as a function of $\alpha$-W fraction, seen in Figure 2c, was extracted using second harmonic Hall measurements on 14 nm-thick W devices that have been uniformly annealed. 2~nm of CoFeB, which has in-plane anisotropy, with a 1~nm MgO/3~nm Ta cap was deposited on top of the W device as the magnetic layer. We employ the standard analysis technique to extract the spin-orbit torques from second harmonic Hall data as described in Ref. \cite{avci2014interplay, ding2024orbital}. The current and magnetic field dependence of the second harmonic Hall signal is fit to extract the damping-like field and associated damping-like torque efficiency. Representative data for the as-grown W/2~nm CoFeB device is shown in Figure \ref{fig:2ndharm}. \np

The angular dependence of the second harmonic Hall resistance, $R_\text{xy}^{2\omega}(\varphi)$, is shown in Figure \ref{fig:2ndharm}a, and the data is fit to the following model \cite{avci2014interplay}:

\begin{equation}
    R_\text{xy}^{2\omega}(\varphi) = \Big(\frac{1}{2}R_\text{AHE}\frac{B_\text{DL}}{\mu_0 H + \mu_0 H_\text{K}^\text{eff}} + R_{\nabla T}\Big)\cos\varphi + R_\text{PHE}\frac{B_\text{FL}+B_\text{Oe}}{\mu_0 H}\Big(2\cos^3\varphi-\cos\varphi\Big)
\end{equation}
where $\varphi$ is the angle between the current direction and the applied magnetic field, $R_\text{AHE}$ is the anomalous Hall resistance, $\mu_0 H$ is the applied magnetic field, $\mu_0 H_\text{K}^\text{eff}$ is the effective anisotropy field, $R_{\nabla T}$ is the anomalous Nernst resistance, $B_\text{DL}$ and $B_\text{FL}$ are the damping-like and field-like fields, and $B_\text{Oe}$ is the Oersted field contribution. \np

For each device, $R_\text{AHE}$ is extracted with a separate measurement of the Hall resistance $R_\text{H}$ as a function of applied out-of-plane magnetic field $\mu_0 H_z$, seen in Figure \ref{fig:2ndharm}b. The difference between the two dashed lines in Figure \ref{fig:2ndharm}b, which are linear fits to the high-field region of $R_\text{H}$, gives $2R_\text{AHE}$, and the field where $R_\text{H}$ reaches this saturated value is defined as $\mu_0 H_\text{K}^\text{eff}$. The magnitude of the $\cos\varphi$ term scaled by $R_\text{AHE}$ is plotted as a function of $(\mu_0 H + \mu_0 H_\text{K}^\text{eff})^{-1}$ in Figure \ref{fig:2ndharm}c for five different currents. The slope of this linear fit is proportional to $B_\text{DL}$, which is plotted as a function of the current flowing through W ($I_\text{W}$) calculated using a parallel resistor model, in Figure \ref{fig:2ndharm}d. $B_\text{FL}+B_\text{Oe}$ is obtained with the same procedure (plotting the magnitude of the $2\cos^3\varphi-\cos\varphi$ as a function of $H^{-1}$). We estimate the magnitude of the Oersted field from $B_\text{Oe}\approx \mu_0 I_\text{W}/2w_\text{HB}$ where $w_\text{HB}$ is the 10~\textmu m Hall bar width. Finally, the slope of the linear relationship between $B_\text{DL(FL)}$ and $I_\text{W}$ is directly proportional to the damping-like (field-like) torque efficiency $\xi_\text{DL(FL)}^j$ through the following relation:

\begin{equation}
    \xi_\text{DL(FL)}^j = \frac{2e}{\hbar}\frac{M_s t_\text{CFB}B_\text{DL(FL)}A}{I_\text{W}}
\end{equation}
\\
where $M_s$ is the saturation magnetization of CoFeB, $t_\text{CFB}$ is the thickness of CoFeB, and $A$ is the cross-sectional area of the device. $\xi_\text{FL}^j$ is shown in Figure~\ref{fig:2ndharm}e. For $\beta$-W, $\xi_\text{FL}^j\approx0$. Devices with a small fraction of $\alpha$-W appear to have a significant field-like torque, but this may be due to a current distribution in the device where the Oersted field approximation fails. Thus, for micromagnetic simulations of W/CoFeB, $\xi_\text{FL}^j$ is set to 0. 

\clearpage

\section{Spin-orbit torque switching}
\begin{figure}[ht!]
    \centering
    \includegraphics[width=0.95\linewidth]{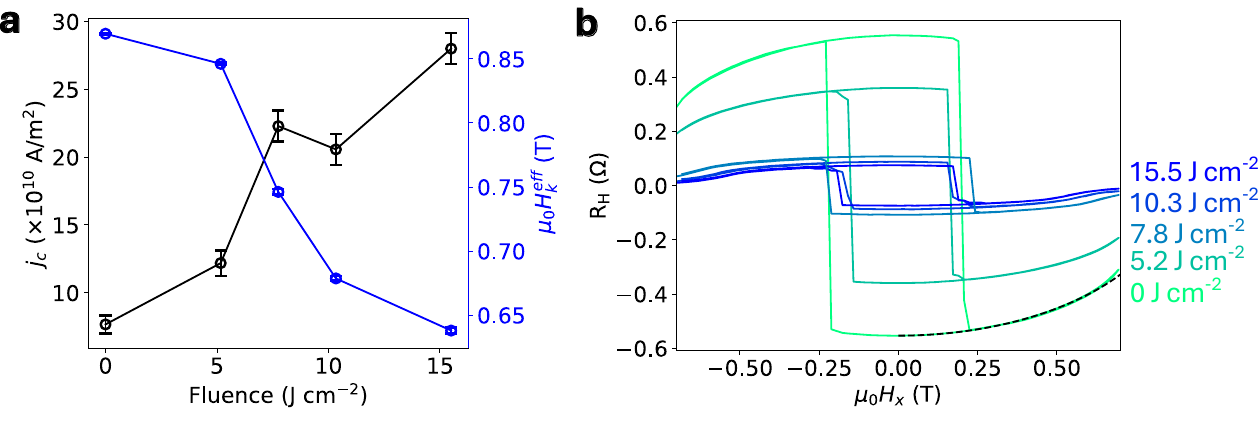}
    \caption{a) The critical switching current density $j_c$ for devices uniformly annealed at different laser fluences with an applied magnetic field of 20 mT is plotted in black. The effective anisotropy strength $\mu_0 H_\text{K}^\text{eff}$ of the CoFeB layer for each device is shown in blue. b) The change in anomalous Hall resistance as a function of in-plane applied magnetic field $\mu_0H_x$ for W/CoFeB devices annealed at different laser fluences. The fit to Equation 3 for the $\beta$-W/CoFeB device is indicated by the black dashed line. }
    \label{fig:switchingcurrent}
\end{figure}

The critical switching current $j_c$ of out-of-plane magnetized W/0.9 nm CoFeB devices with the W layer uniformly annealed in each device at a different laser fluence was determined by tracking the changes in $R_\text{H}$ with current. As the laser fluence, and thus the fraction of $\alpha$-W, was increased, $j_c$ also increases significantly (black data in Figure~\ref{fig:switchingcurrent}a). A decrease in $\mu_0 H_\text{K}^\text{eff}$ of the CoFeB layer is also observed due to a decrease of perpendicular magnetic anisotropy (shown in blue). 

The effective anisotropy strength of CoFeB on different W devices shown in Figure~\ref{fig:switchingcurrent}a is extracted with planar Hall effect measurements, seen in Figure~\ref{fig:switchingcurrent}b. The change in $R_\text{H}$ as a function of $\mu_0H_x$ is fit to the following equation: 
\begin{equation}
    R_H = R_0\sqrt{1-\Big(\frac{\mu_0 H_\text{x}}{\mu_0 H_\text{K}^\text{eff}}\Big)^2}
\end{equation}
where R$_0$ is a scaling factor, and $\mu_0 H_\text{K}^\text{eff}$ is given by $\mu_0 H_\text{K}^\text{eff}$ = $\mu_0$($M_\text{s}+H_\text{K}^\perp$) where \textit{H}$_\text{K}^\perp$ is the perpendicular anisotropy field. $\mu_0 H_\text{K}^\text{eff}$ is defined as positive (negative) when the film has perpendicular (in-plane) anisotropy. An example of the fit is indicated by the black dashed line in Figure~\ref{fig:switchingcurrent}b. 
\clearpage

\section{Finite element and micromagnetic simulations}
\begin{figure}[h!]
    \centering
    \includegraphics[width=0.95\linewidth]{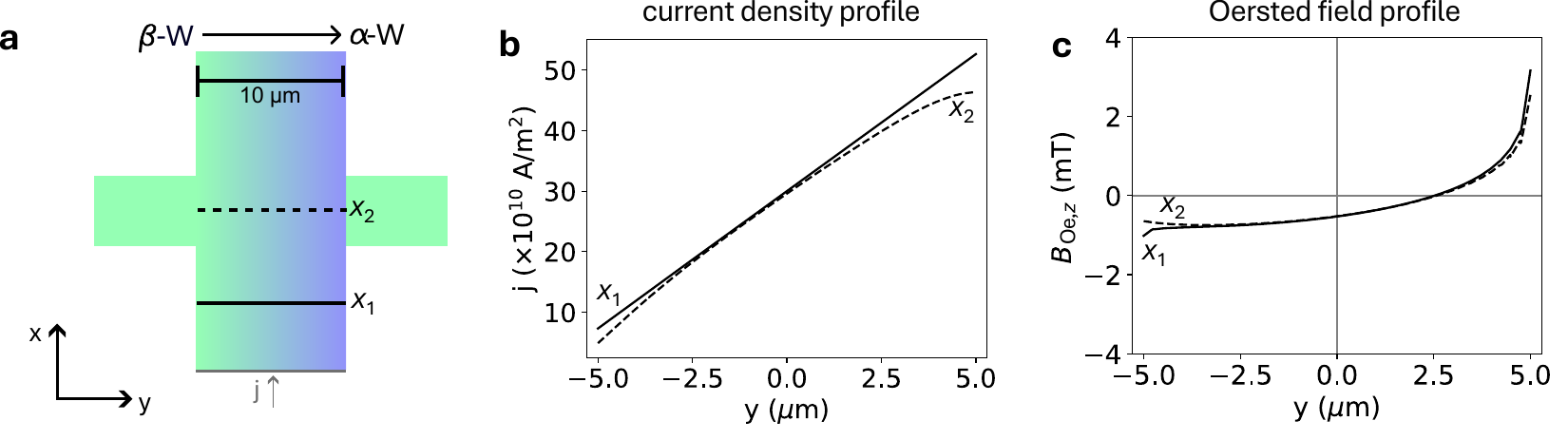}
    \caption{a) Schematic of device geometry for COMSOL simulations. A linear gradient in $\beta$-W going from 0\% to 100\% $\alpha$-W across a 10 \textmu m-wide, 6 nm-thick W wire is simulated, and the current density and Oersted field are evaluated at two positions on the Hall cross: $x_1$, where the device is 10 \textmu m wide (black solid line),  and $x_2$, at the location of the wires for Hall voltage detection (black dashed line). b) $j$ across the device for $x_1$ (black solid line) and $x_2$ (black dashed line). c) The $z$-component of the Oersted field $B_z$ for the current density shown in (b). }
    \label{fig:comsol}
\end{figure}

COMSOL Multiphysics$^{\circledR}$ software \cite{comsol} is used to simulate the current density and Oersted field of W devices with phase gradients. A 6 nm-thick W film with a gradient from 0\% to 100\% $\alpha$-W over 10~\textmu m was simulated, which is much steeper than the 0\% to 10\% and 0\% to 20\% $\alpha$-W gradients used in experimental devices. This steep gradient was chosen in order to observe the largest possible changes in properties across the device (see schematic in Figure~\ref{fig:comsol}a). The nominal current density $j$ of $3\times10^{11}$~A/m$^2$ (see Figure~\ref{fig:comsol}b), which is slightly higher than the current used for field-free switching experiments, is applied for 50 \textmu s. This was used to simulate the highest possible Oersted field (Figure~\ref{fig:comsol}c) to which the device could be exposed in experiments. The current density and Oersted field are shown at two $x$ positions along the Hall cross: at $x_1$, where the current density flows through a straight region, and at $x_2$, where the wires for Hall voltage detection are placed. This accounts for changes in the current density due to the Hall voltage leads. \np

$B_\text{Oe}$ and $j$ as a function of \textit{y} (geometry given in Figure~\ref{fig:comsol}a) are input into a MuMax3 \cite{vansteenkiste2014design} simulation to determine if these conditions are sufficient for field-free magnetization reversal. Changes in CoFeB anisotropy across the laser fluence gradient, seen in Figure~\ref{fig:switchingcurrent}, are also included. A magnetic simulation volume of 250~nm$\times$400~nm$\times$1~nm ($x\times y\times z$) divided into 2 nm cells is implemented with $M_\text{s}$ = 700 kA/m and the exchange energy A$_\text{ex}$=$1\times10^{-11}$~J/m. Unless otherwise stated, the Gilbert damping $\alpha$ is set to 0.20. 50 regions with length along \textit{x} of 250~nm and width along \textit{y} of 8 nm are defined, and each region has the associated current density, Oersted field, anisotropy energy, and spin-orbit torque efficiency corresponding to the phase gradient going from $\beta$-W to $\alpha$-W. Because a module for spin-orbit torques is not available, they are modeled using the spin transfer torque module. This is possible because a current along the \textit{z}-direction for spin transfer torques is equivalent to a current along the \textit{x}-direction for spin-orbit torques. The orientation of the polarized spins is set by the magnetization orientation of the magnetic ``fixed layer" in the spin transfer torque module, and only the damping-like torque is considered (the field-like torque is set to zero with the ``epsilon prime" parameter). After the system is relaxed for 0.2 ns, positive $j$, with densities scaled to account for the current distributions shown in Figure~\ref{fig:comsol}b, are applied for 1 ns, and the non-uniform $B_\text{Oe}$ associated with the current shown in Figure~\ref{fig:comsol}c is also included. Then, $j$ and $B_\text{Oe}$ are removed, and the system is allowed to relax for 5 ns. The out-of-plane magnetization $m_z$ is recorded as a function of time during this process. \np

\begin{figure}[ht!]
    \centering
    \includegraphics[width=0.8\linewidth]{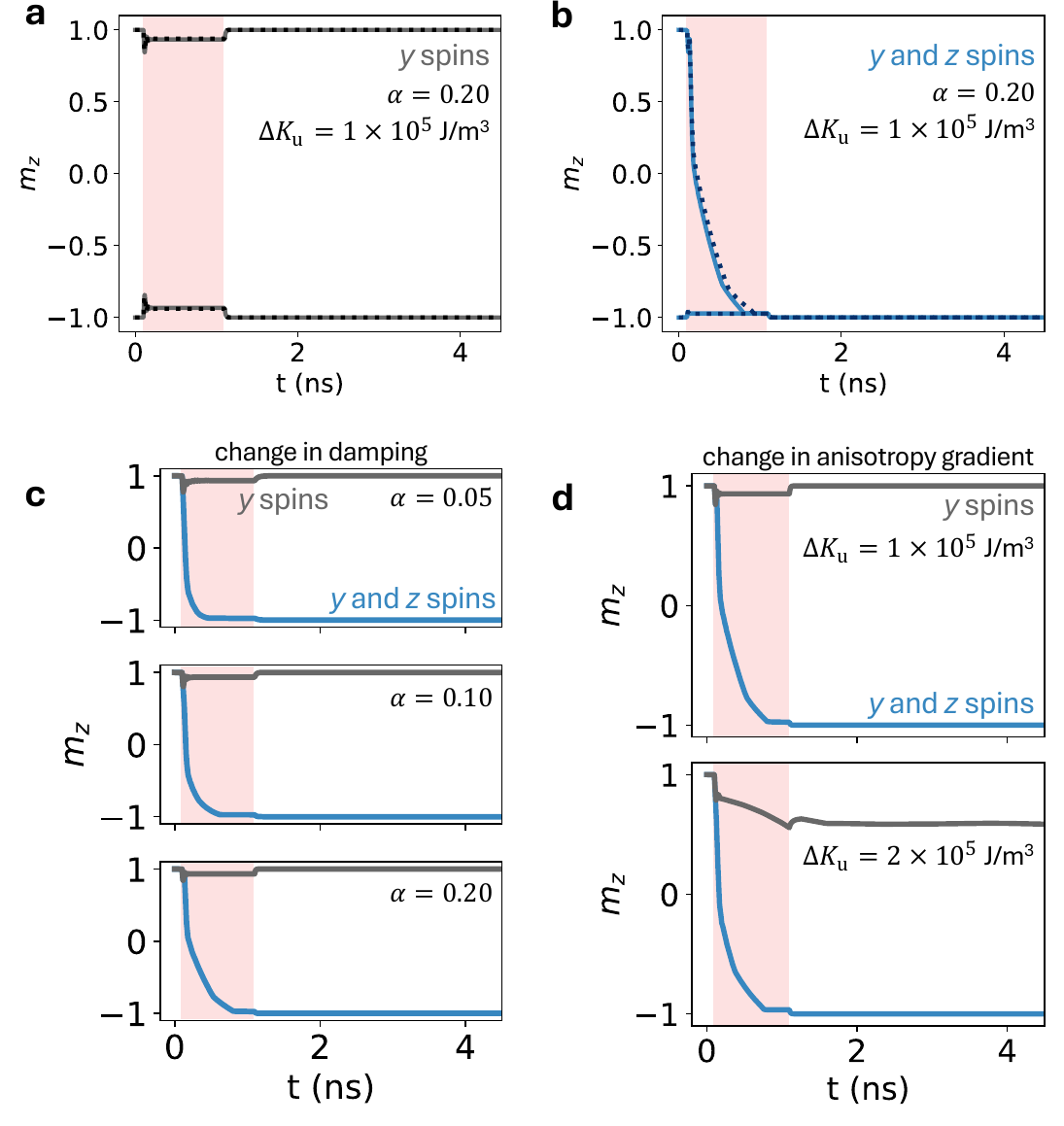}
    \caption{Time traces of $m_z$ before, during, and after application of a 1 ns square current pulse (applied at t=0.2~ns) obtained from micromagnetic simulations including both $j$ and $B_\text{Oe}$. The shaded pink region indicates the time period where the current pulse is applied. The solid (dashed) lines are simulated at position $x_1$ ($x_2$). $j$ and $B_\text{Oe}$ for each position are taken from Figure~\ref{fig:comsol}b and \ref{fig:comsol}c, respectively. a) Response of $m_z$ to spin-orbit torques starting from $m_z=+1$ and $m_z=-1$, assuming only $y$-polarized spins, where no magnetization reversal is observed. b) Response of $m_z$ to spin-orbit torques starting from $m_z=+1$ and $m_z=-1$, assuming $y$ and $z$-polarized spins, where deterministic magnetization reversal is only observed when the initial state of $m_z=+1$. c) The magnetization reversal for $y$-polarized spins (in grey) and $y$ and $z$-polarized spins (in blue) for different values of $\alpha$ ($\Delta K_\text{u}$~= 1.5$\times 10^5$ J/m$^3$) at $x_1$ starting at $m_z=+1$. For all $\alpha$ values, switching is only observed when $z$-polarized spins are included. d) The magnetization reversal for two scenarios: $y$-polarized spins (in grey) and $y$ and $z$-polarized spins (in blue) for different $\Delta K_\text{u}$ across the device ($\alpha=0.20$) at $x_1$ starting at $m_z=+1$. Full magnetization reversal is only seen when $z$-polarized spins are included (in blue).}
    \label{fig:mumaxSims}
\end{figure}

When the spin polarization was set so that there are only conventional \textit{y}-polarized spins, the combined changes in current density, Oersted field, magnetic anisotropy, and spin-orbit torques across the device were not sufficient to switch the magnetization at both location $x_1$ (solid line in Figure~\ref{fig:mumaxSims}a) and $x_2$ (dashed line in Figure~\ref{fig:mumaxSims}a). Complete magnetization reversal from $m_z=+1$ to $m_z=-1$ was only achieved when \textit{z}-polarized spins were added to the simulation, seen in Figure~\ref{fig:mumaxSims}b (at $x_1$ with the solid line and at $x_2$ with the dashed line). When the magnetization is initialized to the $m_z=-1$ state, the current does not switch the magnetization, as expected for deterministic reversal. \np

To check the robustness of the micromagnetic simulations, the same simulation is carried out for three different Gilbert damping ($\alpha$) values. The magnetization dynamics at $x_1$ and $x_2$ are nearly identical (Figure~\ref{fig:mumaxSims}a,b), so additional simulations were only carried out at $x_1$. Regardless of the damping value used, deterministic magnetization reversal is only observed when $z$-polarized spins are included (blue lines in Figure~\ref{fig:mumaxSims}c). Finally, the impact of a thermal gradient is estimated by doubling the anisotropy gradient across the device from $\Delta K_\text{u}$~=(8.5 to 7.5)$\times 10^5$ J/m$^3$ to $\Delta K_\text{u}$~=(8.5 to 6.5)$\times 10^5$ J/m$^3$. When only $y$-polarized spins are used, the magnetization relaxes to a multidomain state, seen by the fact that $m_z$ plateaus at $\sim0.6$ (grey line in Figure~\ref{fig:mumaxSims}d). When $z$-polarized spins are included, deterministic switching is still achieved (blue lines in Figure~\ref{fig:mumaxSims}d). \np

\end{document}